\documentclass[%
 reprint,
 superscriptaddress,
nobibnotes,
 amsmath,amssymb,
 aps,
 prl,
floatfix,
]{revtex4-2}

\usepackage[breaklinks=true]{hyperref}
\usepackage{xcolor}
\usepackage{soul}
\usepackage{graphicx}
\usepackage{svg}
\usepackage{dcolumn}
\usepackage{bm}

\usepackage[normalem]{ulem}
\newcommand{\nuC}{\nu}
\newcommand{\VG}
{V_\mathrm{G}}
\newcommand{\Peh}{P_\textrm{eh}}

\begin{document}

\title{
Same-spin Andreev reflections in the quantum Hall regime: the role of loss}

\author{Chun-Chia Chen}
\email{Corresponding author. Email: chunchia.chen@duke.edu}
\author{Jordan T. McCourt}%

\author{John Chiles}%

\author{Lingfei Zhao}%

\affiliation{%
 Department of Physics, Duke University, Durham, NC 27708, USA
}%

\author{Kenji Watanabe}

\author{Takashi Taniguchi}

\affiliation{
 National Institute for Materials Science, 1-1 Namiki, Tsukuba 305-0044, Japan
}%

\author{François Amet}

\affiliation{
 Department of Physics and Astronomy, Appalachian State University, Boone, NC 28607, USA
}%

\author{Antonio L. R. Manesco}

\affiliation{
 Center for Quantum Devices, Niels Bohr Institute, University of Copenhagen, DK-2100 Copenhagen, Denmark
}%

\author{Harold U. Baranger}

\author{Gleb Finkelstein}
\email{Corresponding author. Email: gleb.finkelstein@duke.edu}
\affiliation{%
 Department of Physics, Duke University, Durham, NC 27708, USA
}%

\date{\today}

\begin{abstract}
The interfaces of superconductors and topological materials hold promise for realizing exotic states and excitations. An important example is provided by the chiral Andreev edge states (CAES), which are formed at interfaces between quantum Hall (QH) states and superconductors (SC). CAES combine electron and hole amplitudes which are hybridized via Andreev reflections. This study explores the spin properties of the CAES through selective spin filtering of the QH edge channels. 
We find robust evidence of spin-flips  accompanying the Andreev processes:
electrons can be reflected from the superconductor as holes in the same spin channel. We demonstrate that the distribution of the reflection probabilities is exponential and then use random matrix theory  
to account for this observation. Finally, we  observe Andreev reflections in the spin-polarized $\nu=1$ case, which is enabled by particle loss. Our findings shed light on the mechanism underlying Andreev reflections of  spin-polarized chiral states. They also demonstrate the importance of considering non-Hermiticity when constructing topological superconductors in hybrid materials.

\end{abstract}

\maketitle

\section{\label{sec:level1}Introduction}
 
Over the years, significant research effort has focused on the search for topological superconductors with the goal of achieving fault-tolerant quantum computation~\cite{Nayak2008}. One compelling platform for engineering these systems uses devices made of a quantum Hall (QH) states in contact with superconductors (SC)~\cite{Qi2011,Mong2014}. Multiple experimental studies on these hybrid structures have been reported over the past few years~\cite{Amet2016,Lee2017,Seredinski2019,Zhao2020,Gul2022,Hatefipour2022,Vignaud2023,Zhao2023,Akiho2024,Zhao2024}.
In particular, our group has previously detected and explored the properties of the chiral Andreev edge states (CAES) -- the QH edge states proximitized by the superconductor ~\cite{Zhao2020,Zhao2023,Zhao2024}. Motivated by growing interest in these states, a number of recent theoretical works have examined their properties~\cite{Manesco2022,Tang2022,Kurilovich2023,David2023,Schiller2023,Michelsen2023,Hu2024,Bondarev2025,AlexeyB_GrS-Nano_PRB26,Parfenov2026,Maji2026}. 
Specifically, it has been identified that the CAES are significantly influenced by the presence of disorder and vortices in the superconductor~\cite{Manesco2022,Kurilovich2023}. 

A CAES is formed when an electron e.g.\ in the spin-up QH channel is Andreev-reflected from the superconductor and turns into a hole in the spin-down channel~\cite{Hoppe2000}. The latter is subsequently reflected as a spin-up electron and the process is repeated to form a CAES made of a linear superposition of a spin-up electron and and a hole in the spin-down channel. At the same time, the spin-down electrons and holes in the spin-up channel are expected to form a separate CAES. Because this picture requires both spin channels, one might expect to find no Andreev reflections at the spin-polarized filling factor $\nu = 1$ unless spin-orbit coupling is involved. Surprisingly, it turns out that
the coupling of an s-wave  superconductor to the $\nu = 1$ QH state is expected to vanish at zero energy even in the presence of spin-orbit coupling~\cite{Fisher1994,beri_dephasing-enabled_2009,vanOstaay2011,S2by2}.

So far, our 
work on CAES focused on the spin-un\-polar\-ized filling factor $\nu =2$ ~\cite{Zhao2020,Zhao2023,Zhao2024}. 
In this study, we use local electrostatic gates to form $\nu= 1$ ``spin-polarizers'' up- and down-stream from the superconductor, which allows us to selectively filter incident and outgoing channels and to study the spin properties of CAES.
When the QH filling factor next to the superconductor is $\nu= 2$, we clearly identify the spin-flip processes, which enable both the normal electron scattering between the two spin channels, and Andreev reflections that result in the appearance of holes in the same-spin channel.

We then demonstrate that CAES can form even when the QH bulk filling near the superconductor is $\nu= 1$, as evidenced by the emergence of holes in the downstream signal. We argue that same-spin Andreev reflections are enabled by the absorption of electrons into the superconducting contact, which violates the unitarity that would otherwise preclude spin flip at $\nu= 1$~\cite{beri_dephasing-enabled_2009,vanOstaay2011,S2by2}. 

Overall, this work advances our understanding of the spin properties of CAES and shows how spin-flip and loss in the superconductor can create independent particle and hole excitations in the outgoing QH edge channel. Furthermore, it presents a concrete example in which non-hermitian physics 
\cite{Ashida-NonHermitianRev-20} is directly relevant for the mesoscopic electronic transport.

\section{Results}

\begin{figure*}
\centering
\includegraphics[width=0.8\textwidth]{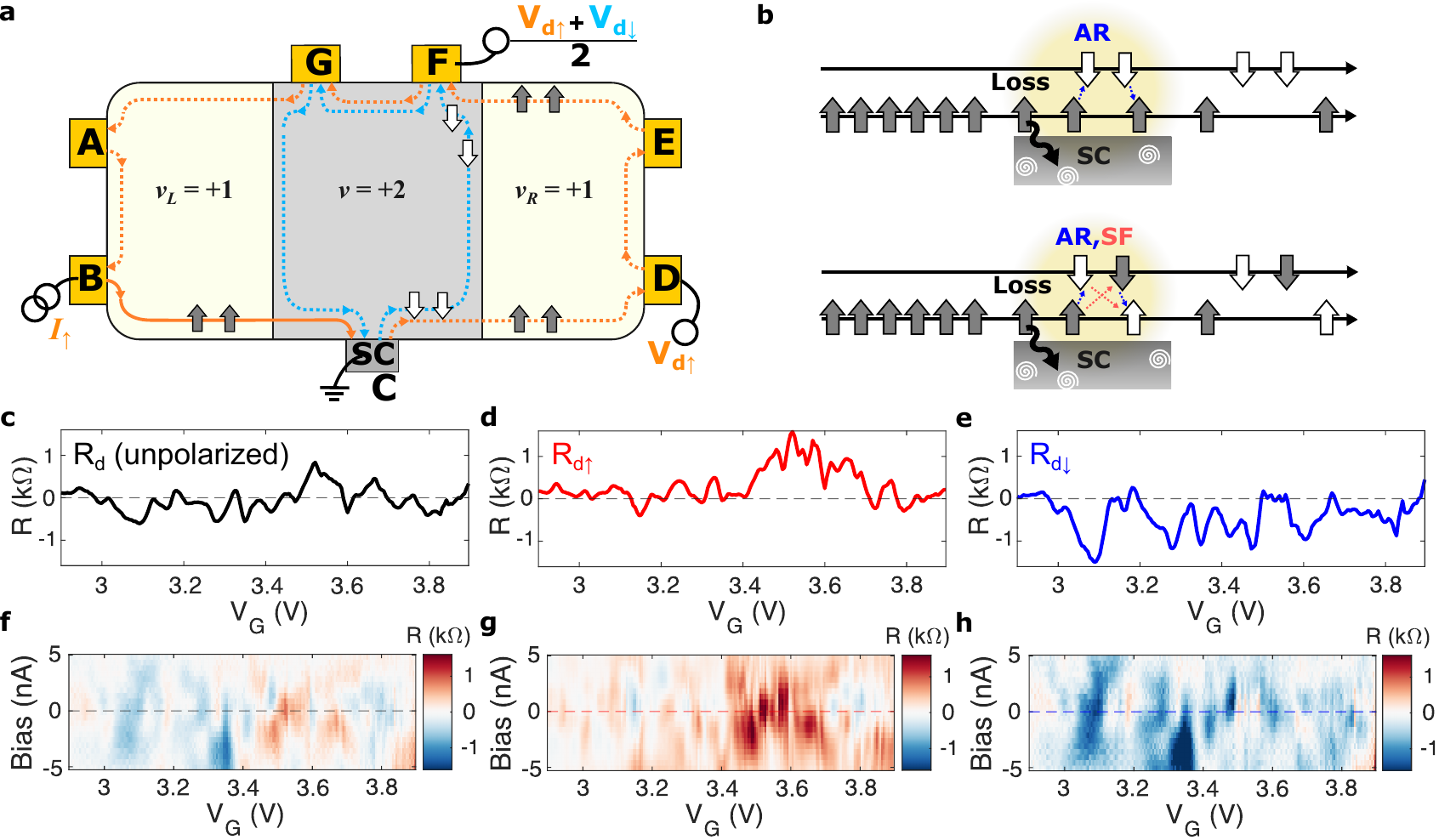}
\caption{\label{fig:spinflip}
a) Schematic of the sample. The left and right regions are kept at $\nu_\mathrm{L,R}=1$, thus forming spin polarizers, while the central region next to the superconducting contact C is separately gated to $\nuC=2$. We apply current bias from contacts B to the grounded superconductor C thereby injecting $e_\uparrow$ to the central region. We measure the downstream voltage at contacts D and F, which allows us to extract the downstream resistances for the same and opposite spin channels, $R_\mathrm{d\uparrow}$ and $R_\mathrm{d\downarrow}$.
b) Schematics of microscopic processes at the SC interface. In the top panel, spin is conserved and only Andreev processes are allowed. In the bottom panel, spin flip is present. Gray and white block arrows indicate electrons and holes. Spirals in the superconductor are vortices, which can absorb quasiparticles.
 c)  Downstream resistance $R_\mathrm{d}$ as measured at contact F as a function of $V_G$
at $B=2.0$ T. The unpolarized resistance
shows random (but reproducible) fluctuations. These are about equally spread between positive and negative values, corresponding to comparable probability of normal and Andreev reflections.
d,e) In contrast, the
$R_\mathrm{d\uparrow}$ and $R_\mathrm{d\downarrow}$ signals are mostly positive or negative, as expected for the predominant emission of $e_\uparrow$ and $h_\downarrow$. However, the negative regions in (d) and positive regions in (e) are clearly visible, indicating
that the $h_\uparrow$ and $e_\downarrow$ are also present. This observation directly proves the existence of spin-flip processes.
f-h) Maps of the unpolarized $R_\mathrm{d}$, $R_\mathrm{d\uparrow}$, and $R_\mathrm{d\downarrow}$ measured as a function of current bias and gate voltage. Panels (c-e) correspond to the zero-bias cuts of these maps. It is clear that the observed behavior is not specific to zero bias and extends over a range of energies. 
}
\end{figure*}

In the main text, we focus on 
the sample schematically shown in Figure~\ref{fig:spinflip}a. The contacts are made of MoRe alloy, a type II superconductor with a critical temperature of $\sim\!8$-$9\,$K and a critical field exceeding 10~T. We study the contact at the bottom center labeled C. It has an interface of length $\sim\!500\,$nm, allowing a substantial fraction of incoming electrons to be reflected to the downstream channel as either electrons or holes~\cite{Zhao2023}. The rest of the contacts have much longer interfaces, so that they effectively serve as normal contacts -- all incoming electrons are fully equilibrated. The left and right regions of the sample are controlled by a graphite back gate. Throughout the paper, they stay fixed at filling factor $\nu_\mathrm{L,R}=1$, thereby serving as spin polarizers, allowing only the spin-up electrons $e_{\uparrow}$ to pass through. Similar sample geometry has been recently proposed in Ref.~\cite{Parfenov2026}. The filling factor $\nuC$ in the central region adjacent to the superconductor is separately controlled by voltage $V_\mathrm{G}$ applied to the Si gate. Except for the last figure, we keep this region on the $\nuC=2$ plateau, resulting in the presence of both spin channels.

To characterize the CAES, we apply bias current $I$ between contact B and the grounded superconductor C, thereby injecting $e_{\uparrow}$ at the superconducting interface. 
We measure the ``downstream'' voltages $V_\mathrm{d}$ at contacts D and F and present these signals as the downstream resistance, $R_\mathrm{d}=dV_\mathrm{d}/dI$. 
Since only the spin-up channel reaches contact D, its voltage directly corresponds to the same spin resistance, $R_\mathrm{d\uparrow}$. The signal at contact F has contribution of both spin channels and therefore represents the unpolarized $R_\mathrm{d}$. (See orange and blue dashed lines in Figure~\ref{fig:spinflip}a, which connect C and F. Note that contacts D and E are voltage probes, which do not drain the spin-up channel.) 
Finally, in order to reveal the spin-flipped signal, $R_\mathrm{d\downarrow}$, we subtract the contribution of the spin-up channel from the voltage measured by contact F.  This procedure has been checked by grounding contact E to drain the spin-up current. We have also verified it in a separate sample, which allowed us to measure both spin channels directly, see Supplementary for further details.

\begin{figure*}
\centering
\includegraphics[width=\textwidth]{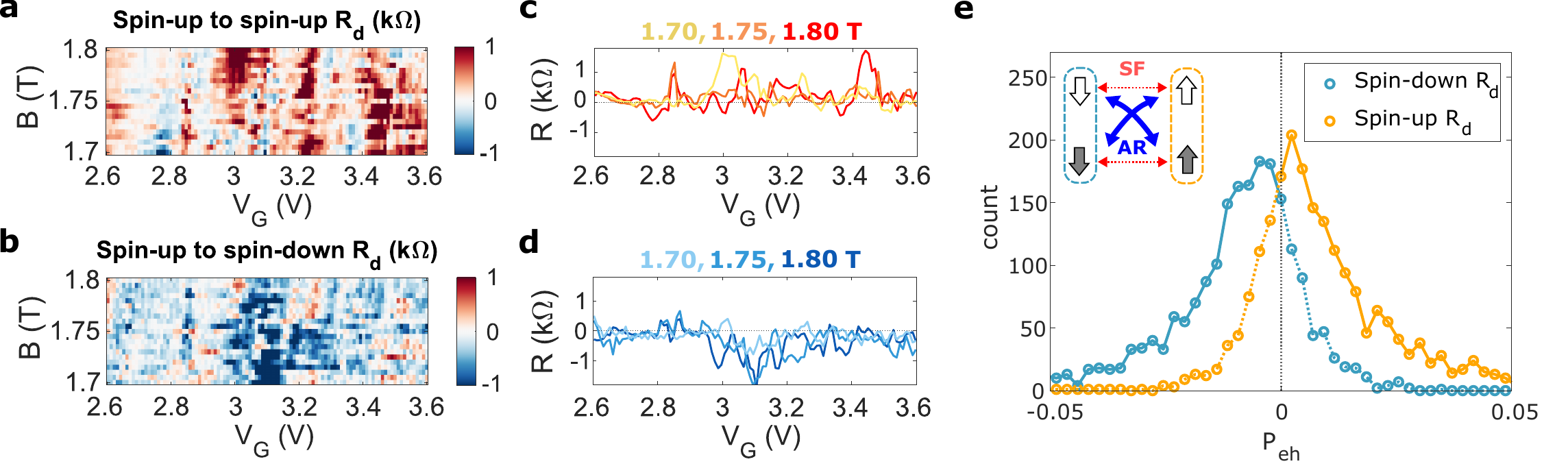}
\caption{\label{fig:multipaths}a,b) $R_\mathrm{d}$ maps as a function of the filling of the central region ($\VG$) and magnetic field for spin-up (same spin) and spin-down (opposite spin) downstream channels.  As in Figure~\ref{fig:spinflip}a, we inject only spin-up electrons at contact B. c,d) Selected horizontal cross-sections of the maps at $B=$1.70~T, 1.75~T, and 1.80~T. e) $P_\mathrm{eh}$ histogram of the data in (a,b). Here, $P_\mathrm{eh}$ is the difference between probabilities to emit as an electron or a hole (see  eq.~\ref{eq:def-Peh}).  The orange and blue curves correspond to the spin-up and spin-down channels. Positive $P_\mathrm{eh}$ indicates the outgoing particles are electrons, and negative values indicate holes. 
Inset in (e) schematically shows that the strong Andreev reflections (AR) yield spin-up and spin-down distributions which are related by reflection around the vertical axis. Note that the blue curve extends to positive $P_\mathrm{eh}$, which corresponds to the normal reflection acompanied by the spin flip. Similarly, the orange curve extends to negative $P_\mathrm{eh}$, which corresponds to Andreev reflections in the same spin channel.}
\end{figure*} 

We start with the unpolarized $R_\mathrm{d}$, Figure~\ref{fig:spinflip}c. 
In a typical measurement, the patterns of $R_\mathrm{d}$ vs $V_\mathrm{G}$ demonstrate random fluctuations, which are reproducible upon successive gate voltage sweeps~\cite{Zhao2020,Zhao2023}. These features extend over a range of energies, as seen in the bias-$V_\mathrm{G}$ map shown in Figure~\ref{fig:spinflip}f.
The fluctuations reflect the disordered nature of the QH-SC interfacial region~\cite{Manesco2022} and the effect of the vortices in the superconductor on the CAES propagating along the interface~\cite{Kurilovich2023}. Importantly, the sign of the downstream resistance depends on whether the particles emitted downstream of the superconductor are predominantly electrons (positive $R_\mathrm{d}$) or holes (negative $R_\mathrm{d}$). The magnitude of $R_\mathrm{d}$ is typically a few \% of the quantum Hall resistance. This reduction indicates that most quasiparticles are absorbed by vortices in the superconductor~\cite{Zhao2023}. 

We next analyze the polarization of the particles emitted downstream. Figure~\ref{fig:spinflip}b schematically shows the expected processes at the SC interface if the spin is conserved (top) and in the presence of spin flips (bottom). Andreev reflections from the grounded superconductor turn the incoming $e_\uparrow$  into holes in the spin-down channel, $h_{\downarrow}$. Without the spin flips, spin and charge would be locked, leading to only electrons in the outgoing spin-up channel, and only holes in the spin-down channel. Therefore, spin conservation would result in positive $R_\mathrm{d\uparrow}$ and negative $R_\mathrm{d\downarrow}$.

We demonstrate the typical behavior of $R_\mathrm{d\uparrow}$ and $R_\mathrm{d\downarrow}$ vs $\VG$ in Figure~\ref{fig:spinflip}d,e. As expected, we find mostly positive $R_\mathrm{d\uparrow}$, however, there are interspersed regions of negative signal. Similarly, $R_\mathrm{d\downarrow}$ is mostly negative but clearly contains regions of positive signal. The same behavior is also seen in the bias-$V_\mathrm{G}$ maps of Figure~\ref{fig:spinflip}g,h, indicating that it extends over a range of energies. While the spin is {\it mostly} conserved, the negative $R_\mathrm{d\uparrow}$ and positive $R_\mathrm{d\downarrow}$ clearly indicate the presence of the spin-flip processes, which allow $e_{\downarrow}$ and $h_{\uparrow}$ to be emitted downstream, see the bottom schematics in Figure~\ref{fig:spinflip}b. 

To quantify the strength of spin-flip processes, we will compare the relative populations of electrons and holes in both spin channels. Due to the mesoscopic randomness of the downstream patterns, we follow Ref.~\cite{Zhao2023} and investigate the distribution of the signals rather than individual traces. To generate a statistically significant dataset on the CAES interference patterns, we take the advantage of the presence of vortices in the superconducting contact. Vortex rearrangements induced while sweeping magnetic field produce discontinuous changes in the CAES interference patterns, as observed in the downstream measurements~\cite{Zhao2023}. This means that by measuring a $B$-$\VG$ maps, we can collect statistics from many different CAES interference patterns due to multiple vortex arrangements. 
Such maps for the same and opposite spin channels are  presented in Figure~\ref{fig:multipaths}a,b. We show horizontal cuts of the maps at 1.7~T, 1.75~T, and 1.8~T in Figure~\ref{fig:multipaths}c,d. As expected, we observe a largely positive $R_\mathrm{d\uparrow}$ and a largely negative $R_\mathrm{d\downarrow}$. It is again clear that electrons and holes are predominantly emitted in the spin-up and spin-down channels, respectively. 

\begin{figure*}
\centering
\includegraphics[width=\textwidth]{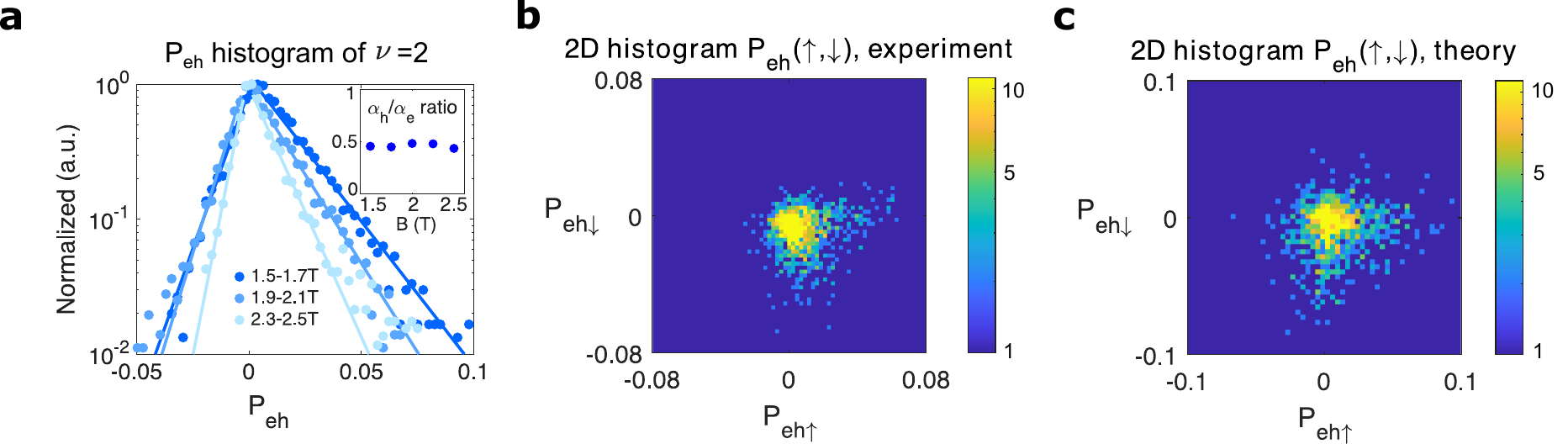}
\caption{\label{fig:RMT}  a) $P_\mathrm{eh}$ histograms similar to Figure~\ref{fig:multipaths}e measured for 3 different magnetic field ranges. The curves are plotted on a logarithmic scale and fitted to an exponential decay on both the electron and hole sides, see Eq.~(\ref{eq:Prob-Peh}). Inset: the ratio of the decay rates $\alpha_\mathrm{h}/ \alpha_\mathrm{e}$ plotted as a function of magnetic field. While both $\alpha_\mathrm{h}$ and $\alpha_\mathrm{e}$ grow with magnetic field, their ratio is found to be nearly constant and close to 0.5. b) 2D histogram $P_\mathrm{eh}(\uparrow,\downarrow)$ with $P_\mathrm{eh}$ values extracted from the simultaneously measured $R_\mathrm{d\uparrow}$ and $R_\mathrm{d\downarrow}$ traces shown in Figure~\ref{fig:multipaths}a,b. Each point of these traces contributes one count to the 2D histogram. 
c) Simulated 2D histogram of $P_\mathrm{eh}$ at $\nuC=2$.  }
\end{figure*}

In Refs.~\cite{Zhao2020,Zhao2024}, we showed that $R_\mathrm{d}$ can be converted to the difference in probabilities of emitting an electron ($P_\mathrm{e}$) or a hole ($P_\mathrm{h}$), namely 
\begin{equation}
    P_\mathrm{eh} \equiv P_\mathrm{e}-P_\mathrm{h} = R_\mathrm{d}/(h/\nu e^2+R_\mathrm{d}) \;.
    \label{eq:def-Peh}
\end{equation}
Due to loss, $R_\mathrm{d}$ is much smaller than the quantum Hall resistance, and  this expression reduces to $P_\mathrm{eh} \approx \nu e^2 R_\mathrm{d}/h$. In the supplementary, we extend this result to the spin-selective case, which allows us to convert the $R_\mathrm{d\uparrow}$ and $R_\mathrm{d\downarrow}$ shown in Figure~\ref{fig:multipaths}a,b to $P_\mathrm{eh\uparrow}$ and $P_\mathrm{eh\downarrow}$. The histograms of these quantities are then plotted in Figure~\ref{fig:multipaths}e. Without spin-flips, the $P_\mathrm{eh\uparrow}$ distribution (orange) would be entirely positive, and likewise, $P_\mathrm{eh\downarrow}$  (blue) would be fully negative. In reality, the presence of spin-flips produces \emph{bipolar} distributions in both spin channels, with a sizable weight of each histogram extending to the opposite side of the $P_\mathrm{eh}$ axis.

We have explored the histograms of spin-unresolved $P_\mathrm{eh}$ in Ref.~\cite{Zhao2023} and have found them to be symmetric around zero. It has been argued that this corresponds to multiple normal and Andreev reflections, which smear the electron/hole population over the phase space~\cite{Kurilovich2023}. The histograms in Figure~\ref{fig:multipaths}e provide a spin-resolved versions of that distribution. Note that the histograms are almost identical when mirrored around zero. Indeed, the formation of the CAES involves multiple reflections, so that the two distributions are converted into each other upon successive Andreev reflections, see schematics in the inset of Figure~\ref{fig:multipaths}e. 

We now focus on the same-spin distribution and plot it for several ranges of magnetic field in Figure~\ref{fig:RMT}a.
Strikingly, we find that the distributions of $P_\mathrm{eh}$ appear linear on a logarithmic scale, both on the electron and the hole sides. This observation holds for several ranges of magnetic field -- while the distribution becomes narrower at higher field, its functional form is preserved.

The exponential behavior of $P_\textrm{eh}$ can be understood within the framework of random matrix theory (RMT)~\cite{Beenakker1997} by considering a large random scattering matrix, $S$. While the number of QH input and output channels here is small, the presence of substantial loss in the system can be modeled by additional effective channels~\cite{ButtikerPRB86, BarangerMello-PhaseBreak-PRB95, BrouwerBeen-PhaseBreak-PRB95},
thus leading to a $S$-matrix of a large rank.
Consideration of a random matrix is justified by the mesoscopic interference patterns observed. For an $S$-matrix of rank $N$, a general result of RMT is that the distribution of a single matrix element $S_{ij}$ in the large $N$ limit is of the form~\cite{PereyraMello-MarginalJPA83, PolianskiBrouwerJPA03, Meckes-RMTbook2019}
\begin{equation}
    \mathcal{P}(|S_{ij}|^2) = \frac{1}{\alpha} e^{-|S_{ij}|^2/\alpha},\quad \alpha \propto \frac{1}{N}\;,
\end{equation}
where the decay rate and prefactor are related by normalization, yielding a single parameter. The exact relation between $\alpha$ and $N$, $\alpha=c/N$, depends on the symmetry of the $S$-matrix: for a pure RMT ensemble $c$ is typically of order $1$, but for a crossover between ensembles, $c$ can be arbitrary and depend on the type of matrix element~\cite{suppinfo}.  

The probabilities for an incoming electron in a single channel to be reflected as an electron or hole in a given outgoing spin-polarized channel
correspond to such $|S_{ij}|^2$. Therefore, their distributions have an exponential form:
\begin{equation}
    \mathcal{P}(P_\mathrm{e}) = \frac{1}{\alpha_\mathrm{e}} e^{-P_\mathrm{e}/\alpha_\mathrm{e}}, \quad
    \mathcal{P}(P_\mathrm{h}) = \frac{1}{\alpha_\mathrm{h}} e^{-P_\mathrm{h}/\alpha_\mathrm{h}} \;.
\end{equation}
We have allowed different decay rates here~\cite{suppinfo} since an electron to hole process in a single channel requires a spin flip from spin-orbit coupling, while an electron to electron process does not.  

The downstream resistance is related to the difference between these two probabilities, Eq.\,(\ref{eq:def-Peh}). Using $\mathcal{P}(P_\mathrm{e})$ and $\mathcal{P}(P_\mathrm{h})$, we find the distribution of $P_\textrm{eh}$ by convolution; a trivial integration yields  
\begin{equation}
    \mathcal{P}(P_\mathrm{eh}) = \frac{1}{\alpha_\mathrm{e} + \alpha_\mathrm{h}}
\begin{cases}
e^{-P_\mathrm{eh}/\alpha_\mathrm{e}}, & P_\mathrm{eh} > 0,\\[6pt]
e^{-|P_\mathrm{eh}|/\alpha_\mathrm{h}}, & P_\mathrm{eh} < 0.
\end{cases}
\label{eq:Prob-Peh}
\end{equation}

This result shows that for a given spin channel, the decay rate on the electron side ($P_\mathrm{eh}>0$) is determined solely by $\alpha_\mathrm{e}$, while the decay rate on the hole side ($P_\mathrm{eh}<0$) is determined by $\alpha_\mathrm{h}$. Since $\alpha_\mathrm{h}$ is the average probability of observing a hole in the outgoing spin-up channel,  $\alpha_\mathrm{h} = \langle P_\mathrm{h} \rangle$, the faster decay on the hole side observed in the spin-up data arises naturally from the smaller number of hole quasiparticles produced by rare spin-flip scattering processes.

\begin{figure*}
\centering
\includegraphics[width=0.8\textwidth]{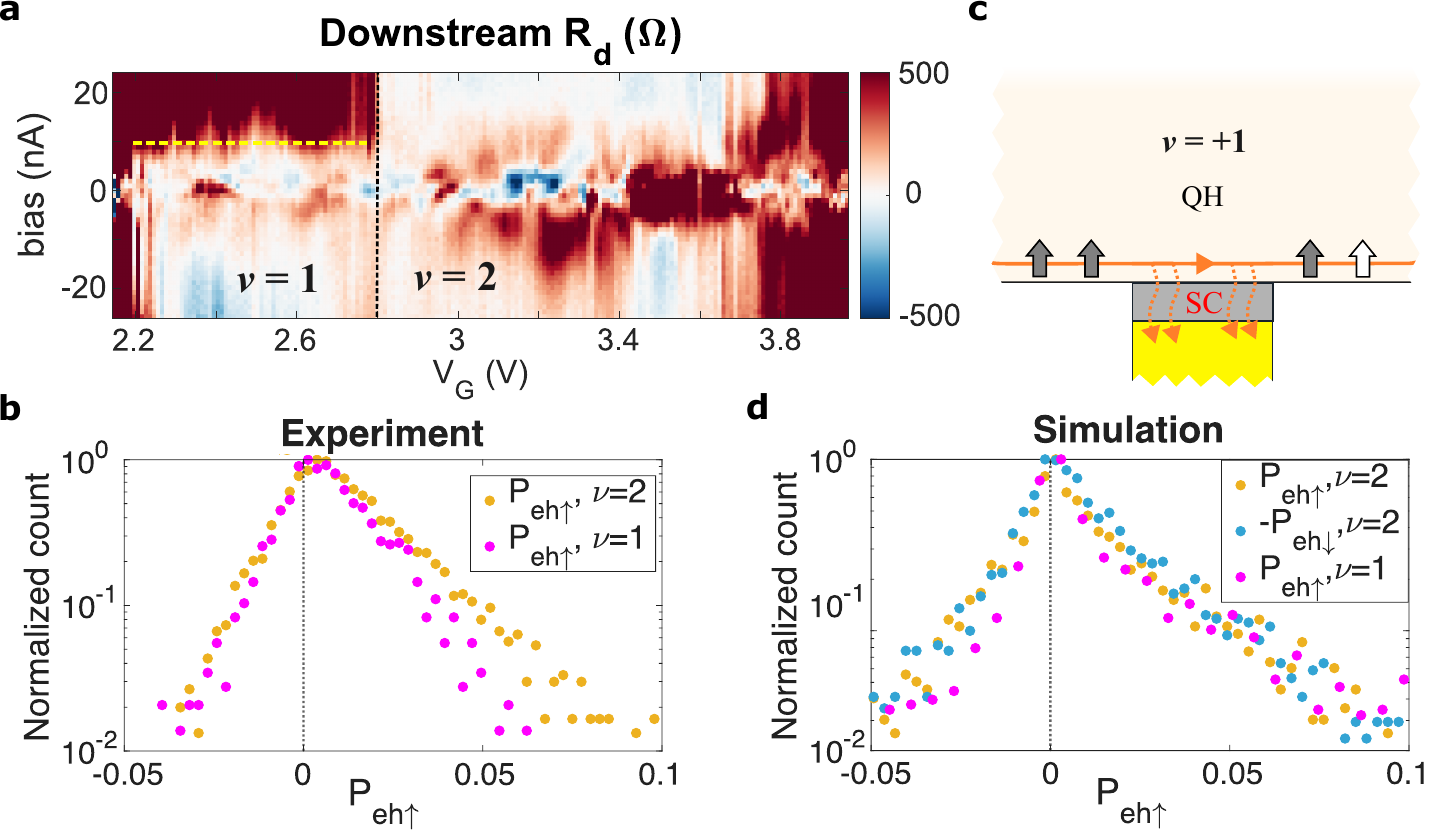}
\caption{\label{fig:nu1} a) Downstream resistance $R_\mathrm{d\uparrow}$ as a function of current bias and $V_\mathrm{G}$ measured at 2~T. The left and right contact regions (Figure~\ref{fig:spinflip}a) are fixed at $\nu_\mathrm{L,R}=1$, while the central region is swept from $\nuC=1$ to $2$. Both parts of the map exhibit qualitatively similar behavior. The yellow dashed line at about 10 nA for $\nuC=1$ indicates the magnon excitation threshold~\cite{Wei2018}. 
Note that for $\nuC=1$ only the spin-up channel can exit the interfacial region. 
b) $P_\mathrm{eh\uparrow}$ histograms for $\nuC=1$ and $\nuC=2$ measured in the $1.5\,$-$1.7$~T range. (Same dataset as one of the magnetic field-dependent histograms in supplementary Figure~S3~\cite{suppinfo}.) c) Schematics of the simulation geometry. The graphene nanoribbon at the top is in the QH regime with $\nuC=1$. The bottom lead is made of a superconducting (gray) and normal (yellow) metal. By making only a part of the lead superconducting, the quasiparticles are allowed to tunnel from the QH ribbon to the normal metal through the superconducting region. The setup keeps the spin-orbit coupling from the superconductor while introducing a loss mechanism. d) $P_\mathrm{eh}$ histograms for $\nuC=1$ and $\nuC=2$ obtained in the simulation. The spin-down data has the $P_\mathrm{eh}$ axis flipped to overlay it with the spin-up case. }
\end{figure*}

We can now quantify the strength of the spin-flip scattering by calculating $\alpha_\mathrm{h}/\alpha_\mathrm{e}$ -- the relative population of electrons and holes in the outgoing spin-up channel. We extract $\alpha_\mathrm{h}/\alpha_\mathrm{e}$ from the histograms at different magnetic-field ranges, as shown in the inset of Fig.~\ref{fig:RMT}a. This ratio remains close to 0.5 and is nearly independent of magnetic field. 
This observation suggests that the spin-flips are determined by local material properties, likely the spin-orbit coupling associated with the heavy metals forming the superconductor. These processes are not expected to depend on magnetic field, even though the downstream signal decreases as vortices proliferate in the superconductor. An alternative explanation would be for the spin flips to originate in the spin-orbit interaction at the interface. However, if this were the case, the ratio $\alpha_\mathrm{h}/\alpha_\mathrm{e}$ would decrease as the electron wavefunction is pushed closer to the interface at higher magnetic field, which is not the case experimentally. 

Beyond the exponential shape of the distributions, the presence of loss has an important consequence for the correlation between the two spin channels. Namely, for strong loss (i.e.\ a large number of effective channels $N$), $P_\mathrm{eh\uparrow}$ and $P_\mathrm{eh\downarrow}$ are expected to be uncorrelated, because they corrspond to different elements of a large random matrix \cite{PereyraMello-MarginalJPA83, PolianskiBrouwerJPA03, Meckes-RMTbook2019}.  In Figure~\ref{fig:RMT}b, we analyze their joint histogram 
$P_\mathrm{eh}(\uparrow,\downarrow)$. Here, each data point corresponds to the pair $(P_\mathrm{eh\uparrow}, P_\mathrm{eh\downarrow})$ extracted from the simultaneously measured traces of $R_\mathrm{d\uparrow}$ and $R_\mathrm{d\downarrow}$. (Same data as in Figure~\ref{fig:multipaths}.)
For uncorrelated channels, the distribution is expected to satisfy
\begin{equation}
\frac{P_\mathrm{eh}(\uparrow,\downarrow)}{P_\mathrm{eh\uparrow} \cdot P_\mathrm{eh\downarrow}} = 1.
\end{equation}
To obtain this denominator, we compute the outer product of the individual one-dimensional (1D) histograms of $P_\mathrm{eh\uparrow}$ and $P_\mathrm{eh\downarrow}$, as shown in supplementary Figure~S4b~\cite{suppinfo}. 
Dividing the original 2D histogram by this uncorrelated reference yields the 2D normalized intensity map shown in Figure~S4c. A plateau with a value close to 1 appears in the central region (Figure~S4d), consistent with the expectation of uncorrelated signals.
As a final piece of evidence, in Figure~\ref{fig:RMT}c we present the joint histogram of the two spin channels produced in a tight binding simulation described later in the text. The simulation confirms that there is little correlation between the two channels, and the similarity with the experiment is very clear.

Having completed the general discussion of the $P_\mathrm{eh}$ distribution, we now return to the holes observed in the same-spin channel. Their presence might be interpreted as indicating that the superconductor can hybridize with a single spin channel---a desired feature for creating chiral Majorana fermions~\cite{Qi2011}.
To explore this possibility, we tune the filling factor in the central region to $\nuC=1$. Theoretically, it is expected that Andreev reflections should be suppressed in this regime, even in the presence of the spin-orbit interaction~\cite{Fisher1994,beri_dephasing-enabled_2009,vanOstaay2011}. Figure~\ref{fig:nu1}a shows that this is not the case. 
Here, we show the $I$-$V_\mathrm{G}$ map of $R_\mathrm{d\uparrow}$, covering both the $\nuC = 1$ and $\nuC = 2$ regions as controlled by $\VG$. For both filling factors, we observe CAES interference patterns dominated by normal reflections (red), but Andreev reflection is also present (blue). 

We further collected statistics in the magnetic field range 1.5 to 1.7~T and extracted histograms of $P_\mathrm{eh\uparrow}$, as in Figure~\ref{fig:RMT}a, for $\nuC = 1$ and 2 -- see Figure~\ref{fig:nu1}b. 
Strikingly, the distributions for $\nuC = 1$ and $\nuC = 2$ are nearly identical. They are both asymmetric and decay exponentially on both the electron and hole side. 

To further understand the appearance of holes in the $\nuC \!=\! 1$ signal and its similarity to the $\nuC \!=\! 2$ case, we have carried out tight-binding simulations using the Kwant quantum transport package~\cite{GrothKwantNJP14}.
We simulate the three-terminal T-shaped geometry with a honeycomb lattice from Ref.~\cite{Manesco2022}, as shown in Figure~\ref{fig:nu1}c.  Here, the QH nanoribbon forms a horizontal interface with the superconducting lead, which has both disorder and spin-orbit coupling.
Controlled loss is introduced by making only a finite segment of the vertical lead superconducting, which allows the particles to tunnel from the QH nanoribbon through the superconductor to the normal part of the lead, Figure~\ref{fig:nu1}c.
Further details of the tight-binding simulations are given in the Supplementary~\cite{suppinfo}.

The distribution of $P_\mathrm{eh}$ from the simulations are shown in Figure~\ref{fig:nu1}d for both $\nuC \!=\! 1$ and $\nuC \!=\! 2$.  Qualitatively, the simulation results are remarkably similar to the experiment: 
(i)~the spin resolved probability is bipolar, even for $\nuC \!=\! 1$, (ii)~the polarity that requires spin flip is less probable, and (iii)~for $\nuC \!=\! 2$, flipping $\mathcal{P}(P_\mathrm{eh\downarrow})$ about the origin yields $\mathcal{P}(P_\mathrm{eh\uparrow})$. As noted for the experiment, the distributions for $\nuC \!=\! 1$ and $\nuC \!=\! 2$ are nearly identical in this regime.

The appearance of holes in the outgoing $\nuC\!=\!1$ channel  can be explained using the following arguments. In contrast to the vacuum edge of the QH region, 
the boundary conditions at the superconducting interface support CAES having both spin directions---$e_\uparrow /h_\downarrow$ and $e_\downarrow / h_\uparrow$ even at $\nuC=1$~\cite{Bondarev2025}. (The two states are conjugate and correspond to the Bogoliubov-de Gennes doubling.)
The incoming $e_\uparrow$ connects to the $e_\uparrow / h_\downarrow$ CAES. 
Spin-flip scattering in the superconductor then couples it to the other CAES, $e_\downarrow / h_\uparrow$ \cite{Kurilovich2023Criticality}, which could then be emitted downstream as $h_\uparrow$. Without loss, particle-hole symmetry together with unitarity prohibit this process in the $\nuC=1$ case~\cite{beri_dephasing-enabled_2009,vanOstaay2011,S2by2}.
However, the prohibition
does not apply in the presence of loss, 
and so we observe spin-up holes in the outgoing channel.

Loss can also explain our second observation -- the similarity between the $\nuC \!=\! 1$ and $\nuC \!=\! 2$ distributions. First, note that the outgoing $e_\uparrow$ and $h_\uparrow$ are independent of each other -- 
they are produced by separate processes in which loss is essential, and their distributions are characterized by different probabilities.
The presence of one extra channel in the $\nuC \!=\! 2$ case vs  $\nuC \!=\! 1$ is then insignificant compared to the number of the lossy channels.

Finally, we note that local doping likely exists near the superconducting contact: the map in Figure~\ref{fig:nu1}a at $\nuC=1$ displays a hint of a magnon emission gap around 10~nA (white dashed line), which indicates the presence of a local filling factor $\nu = 2$ near the contact~\cite{Wei2018}.
The doping may induce a local spin-down electron state in addition to the global spin-up state, making the nominal $\nuC=1$ appear similar to $\nuC=2$. However, in the absence of loss, Andreev reflections producing $h_\uparrow$ in the downstream channel are suppressed for $\nuC=1$ 
because of the same symmetry arguments, even in the presence of doping~\cite{S2by2}. Since this constraint does not exist for $\nuC=2$, in the absence of loss the two cases would look very different.
As explained in the preceding paragraph, loss is essential for the appearance of holes in the downstream channel for $\nuC=1$, making this case so similar to $\nuC=2$. 

\section{Conclusion}
In summary, we have investigated the spin properties of quantum Hall - superconductor interfaces in graphene. We clearly observe spin flip processes, which enable Andreev reflection of electrons to holes in the same spin channel. Our results are supported by a large statistics of CAES interference patterns obtained from unique vortex configurations produced by sweeping the magnetic field. 
From these data, we can also conclude that the scattering into the spin-up and the spin-down channels proceed independently, as evidenced by the 2D histograms. We explain this observation by the strong loss present in our system. Random matrix formalism allows us to account for both the independence of the spin channels and the exponential shape of the distributions in Figure~\ref{fig:RMT}.

Our most striking observation is that for a single spin-polarized QH edge channel, downstream electrons and holes \textit{both} occur. In fact, the distributions of $\Peh$ for $\nuC\!=\!1$ and $\nuC\!=\!2$ are nearly the same.  One may be tempted to infer, incorrectly, that an exotic CAES exists in which electrons and holes in the same spin-polarized channel are hybridized, $e_\uparrow/h_\uparrow$.  Instead, we argue that 
the correct description should be based on the existence of both the $e_\uparrow/h_\downarrow$ and $e_\downarrow/h_\uparrow$ modes ~\cite{Bondarev2025}. In the presence of loss these states are allowed to couple, and the holes in the spin-up channel can then be emitted downstream from the $e_\downarrow/h_\uparrow$ mode. This result has important implications in the ongoing search for 1D chiral Majorana fermions~\cite{Qi2011} and the signature they produce in quantum transport measurements~\cite{DayManesco-loss-SPost25}.

\section{Acknowledgments}
\begin{acknowledgments}
G.F. thanks C. Beenakker, S. Bergeret, L. Glazman, M. Houzet, and D. Loss for valuable theoretical insights. We also thank Ethan Arnault for the discussions of the experimental data. Transport measurements conducted by C.C. were supported by the Division of Materials Sciences and Engineering, Office of Basic Energy Sciences, U.S. Department of Energy, under Award No. DESC0002765. 
Lithographic fabrication and characterization performed by L.Z. and C.C. and experimental data analysis by C.C., J.M., J.C., and G.F. were supported by NSF Award No. DMR-2428579. Theoretical discussion based on the random matrix  approach was developed by H.U.B. Simulations were conducted by A.L.R.M. who acknowledges the funding from the European Research Council (Grant Agreement No.~856526). K.W. and T.T. acknowledge support from the Elemental Strategy Initiative conducted by the MEXT, Japan (Grant No. JPMXP0112101001) and JSPSKAKENHI (Grants No. 19H05790, No. 20H00354, and No. 21H05233).
This work was performed in part at the Duke University Shared Materials Instrumentation Facility (SMIF) (RRID:SCR-027480), a member of the North Carolina Research Triangle Nanotechnology Network (RTNN), which is supported by the National Science Foundation (award number ECCS-2025064) as part of the National Nanotechnology Coordinated Infrastructure (NNCI).
\end{acknowledgments}

\paragraph*{AI usage disclosure}
We have used GitHub Copilot with the GPT-5.2 Codex model and OpenAI Codex with the GPT-5.5 Codex model to set up the numerical simulations.

\medskip
The data that support the findings of this article are openly available~\cite{rigotti_manesco_2026_21440265,Chen_2026_21436565}.

\newpage

\bibliography{Reference,QTransport_2026-07,RMT_gr-S}

\end{document}


\title{Supplementary for\\
``Same-spin Andreev reflection in the quantum Hall regime: \\the role of loss''}

\author
{Chun-Chia Chen*, Jordan T. McCourt$^1$,  John Chiles$^1$, Lingfei Zhao$^1$,\\ Kenji Watanabe$^2$, Takashi Taniguchi$^2$,\\ Fran\c{c}ois Amet$^3$, Antonio L. R. Manesco$^4$,
Harold U. Baranger$^1$, Gleb Finkelstein$^1$
\\[0.3cm]
\normalsize{$^{1}$Department of Physics, Duke University, Durham, NC 27701, USA}\\
\normalsize{$^{2}$National Institute for Materials Science, Tsukuba, 305-0044, Japan}\\
\normalsize{$^3$Department of Physics and Astonomy, Appalachian State University, Boone, NC 28607, USA}\\
\normalsize{$^4$Center for Quantum Devices, Niels Bohr Institute, University of Copenhagen, DK-2100 Copenhagen, Denmark}\\[0.3cm]
\normalsize{$^\ast$To whom correspondence should be addressed; E-mail:  chunchia.chen@duke.edu}}

\maketitle

\tableofcontents

\newpage
\subsection{Sample image and measurement schematic}

\begin{figure*}[htp]
    \centering
    \includegraphics[width=\columnwidth]{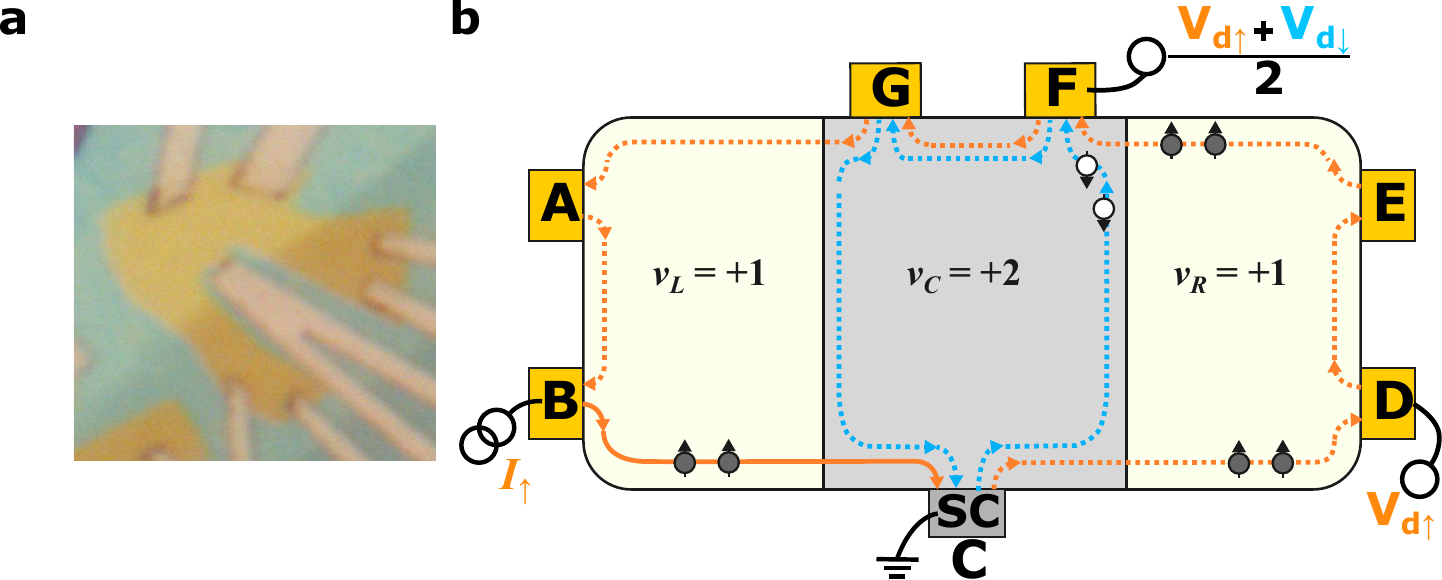}
    \caption {\label{fig:optical}a) Optical image of the device in the main text. The graphene device is encapsulated with top and bottom hBN. Graphite extends partially underneath the stack (darker bottom-right half of the image). The device is etched with ChF$_{3}$O$_{2}$ and SF$_{6}$ plasma for mesa and contacts. The sample is contacted  with sputtered molybdenum-rhenium (MoRe). 
b) Schematic of the device. The orange route is the spin-up QH edge channel. The blue route is the spin-down QH edge channel. The solid orange line indicates that the current is biased from contact B to contact C. The dashed blue and orange lines are  downstream channels reflected from contact C. When dashed lines go in and out of contacts, e.g. E and D, they are not drained since these contacts are not grounded -- the only grounded contact is the SC contact C. As a result, contacts D and E have the same potential. For measuring $V_\mathrm{d\uparrow}$, we probe the voltage difference between contact D and C. For measuring $V_\mathrm{d\downarrow}$, we probe the voltage difference between contact F and C and subtract the contribution of $V_\mathrm{d\uparrow}$. 
    }
\end{figure*}

\subsection{Spin-dependent downstream signal }

\begin{figure*}[h!]
    \centering
    \includegraphics[width=1\textwidth]{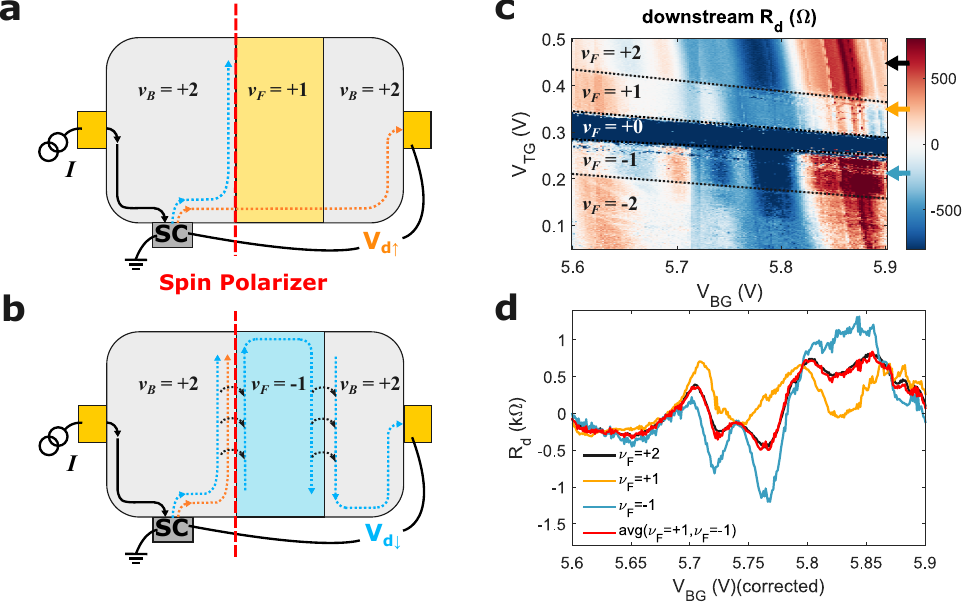}
    \caption {\label{fig:spinfilter} a,b) Schematics of the second device (not the one in the main text) used for exploring the spin filtering. 
    Current bias is applied from the contact on the left to the grounded superconducting contact (bottom left). The non-local downstream voltage potential is measured at the contact on the right. The central region of the sample is defined by a top gate and serves as a spin filter: depending on the filling factor tuned by $V_\mathrm{TG}$, the two spin species can be allowed to pass or be blocked. In panel (a), $\nu_\mathrm{F}=+1$ and only the spin-up channel (orange) is allowed to pass, while the spin-down channel (blue) is blocked. In panel (b), $\nu_\mathrm{F}=-1$ and the role of the spin channels is reversed. The red dashed line in the schematic represents the interface that selects which spin species can pass. c) Downstream resistance map, $R_\mathrm{d}$, measured vs $V_\mathrm{TG}$ and $V_\mathrm{BG}$ at $B=3$ T. Labels next to the y-axis indicate $\nu_\mathrm{F}$, the filling factor under the top gate, viewed as a filter. Across the dashed inclined lines, $\nu_\mathrm{F}$ changes by 1, allowing different spins to pass through the filter.  
    d) Several $R_\mathrm{d}(V_\mathrm{BG})$ curves measured at the $V_\mathrm{TG}$ values indicated with arrows on the map in panel (c). These correspond to $\nu_\mathrm{F}$ equal to $+2$ (black), $+1$ (orange), and $-1$ (blue). To account for the gate cross-talk visible as a slope in (c), the curves are shifted in $V_\mathrm{BG}$ to align their major features. The red curve is the average resistance of $\nu_\mathrm{F}=+1$ and $\nu_\mathrm{F}=-1$ curves. It overlaps with the $\nu_\mathrm{F}=+2$ curve, confirming the proper operation of the spin filter.
    }
    \label{fig:FigS4}
\end{figure*}

In this section, we demonstrate that the downstream signals are spin-dependent. We use a simplified version of the device, which is designed to prove the efficacy of using the spin filter in our problem. The device is a standard encapsulated graphene sample which is divided into three regions (see simplified schematics in Figure~\ref{fig:spinfilter}a and ~\ref{fig:spinfilter}b). The left and right regions are tuned to $\nu_\textrm{B} = 2$ via the bottom Si gate ($V_\textrm{BG}$), and the central region is separately gated by a Pd/Au top gate ($V_\mathrm{TG}$), which we use to tune across $\nu_\mathrm{F} = \pm 1,2$. Here, $\nu_{F}$ denotes the filling factor of the top-gated region viewed as a spin filter. The left region contains a grounded superconducting contact, which is current-biased by an upstream normal metal contact. (Only three contacts are shown for simplicity.) We measure the voltage at the downstream contact located across the spin filter in the right region. The resulting non-local ``downstream resistance'' is defined as $R_\mathrm{d}=dV_\mathrm{d}/dI$ \cite{Lee2017,Zhao2020}. In this sample, we can selectively identify the spin of the particles that pass through the spin filter. 

As depicted in Figure~\ref{fig:spinfilter}a, when a junction is formed between $\nu=+2$ and $\nu=+1$ regions, the spin-up channel passes through the filter, while the spin-down channel is diverted. Similarly, for a junction between $\nu=+2$ and $\nu=-1$ regions (Figure~\ref{fig:spinfilter}b), only the spin-down channel can tunnel through ~\cite{Wei2017}. This results in the switchable circuit where either spin-up (orange lines) or spin-down channel (blue lines) passes through the filter and is measured.

The effect of the spin filter is shown in Figure~\ref{fig:spinfilter}c, where we plot downstream resistance $R_\mathrm{d}$ as a function of top gate voltage $V_\mathrm{TG}$ and back gate voltage $V_\mathrm{BG}$. The latter is swept over a narrow range to ensure that the region interfaced with the superconductor remains on the $\nu=2$ plateau. The patterns of $R_\mathrm{d}$ switch abruptly across the dotted lines overlaid on top of the map. These boundaries correspond to changes of $\nu_\mathrm{F}$ by one, and their slope corresponds to the fact that the central region is influenced by both $V_\mathrm{TG}$ and $V_\mathrm{BG}$.

While the two spin species are equally populated at the source, Figure~\ref{fig:spinfilter}c clearly shows that they undergo different evolution along the superconducting contact. To quantitatively benchmark this effect, in Figure~\ref{fig:spinfilter}d we plot $R_\mathrm{d}$ corresponding to the cuts in Figure~\ref{fig:spinfilter}c. Note that the red and blue features in Figure~\ref{fig:spinfilter}c were not perfectly vertical. This indicates that the electron density near the SC contact is slightly influenced by $V_\mathrm{TG}$. We shift the patterns in Figure~\ref{fig:spinfilter}d to align the features and label the horizontal axis as the ``corrected'' $V_\mathrm{BG}$. 

The blue, orange and black curves in Figure~\ref{fig:spinfilter}d correspond to $\nu_\mathrm{F} = -1,+1,$ and $+2$, respectively. 
Specifically, the black curve corresponds to the regime when all of the sample is at $\nu=+2$ and the downstream CAES is not spin filtered. The $\nu_\mathrm{F} = +1$ case (orange) captures the spin-up component of the downstream signal $R_\mathrm{d\uparrow}$, while the $\nu_\mathrm{F}=-1$ line (blue) represents the spin-down component, $R_\mathrm{d\downarrow}$. The red curve corresponds to the average of these two, $\frac{1}{2}(R_\mathrm{d\uparrow}+R_\mathrm{d\downarrow})$ and clearly overlaps closely with the black curve. This indicates that the average of its spin-polarized components captures the full $R_\mathrm{d}$ pattern.

\subsection{Exponential form of $P_\mathrm{eh}$ histograms 
}
\begin{figure*}[htp]
    \centering
    \includegraphics[width=1\columnwidth]{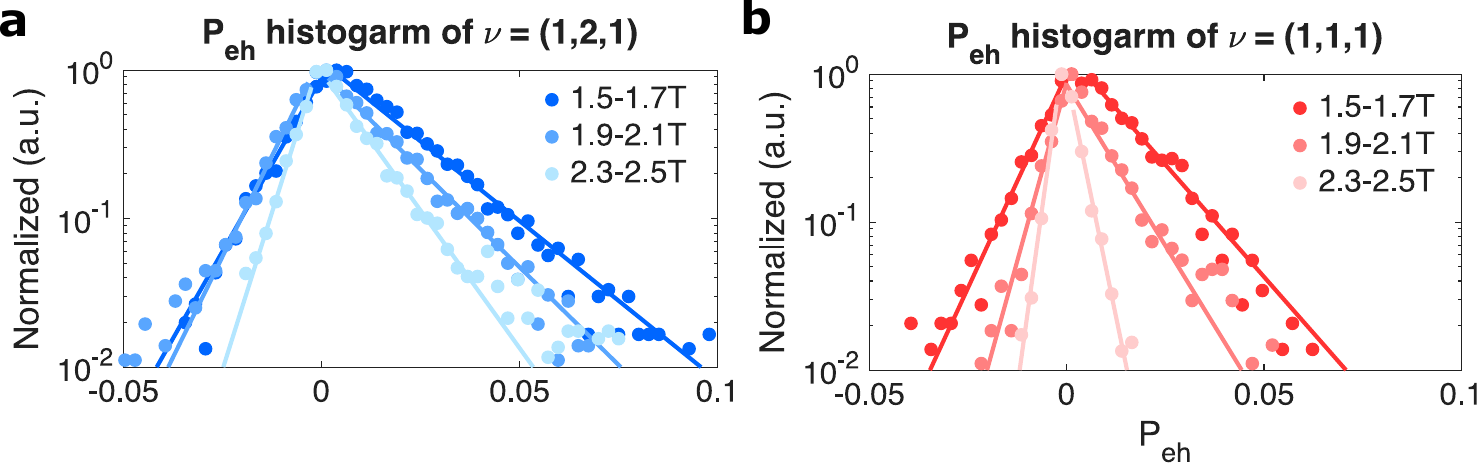}
    \caption {\label{fig:exponential}
    Normalized $P_\mathrm{eh}$ histograms for the sample studied in the main text. Here, panel (a) with $\nu=2$  reproduces Figure~3a. Panel (b) corresponds to $\nu=1$. The filling in the outside regions is fixed at $\nu_\mathrm{L}=\nu_\mathrm{R}=1$. The curves are plotted on a logarithmic scale and fitted to an exponential form on both the electron and hole sides, see main text.
    }
\end{figure*}

\subsection{2D histograms of $P_\mathrm{eh\uparrow}$ and $P_\mathrm{eh\downarrow}$}

\begin{figure*}[htp]
    \centering
    \includegraphics[width=1\columnwidth]{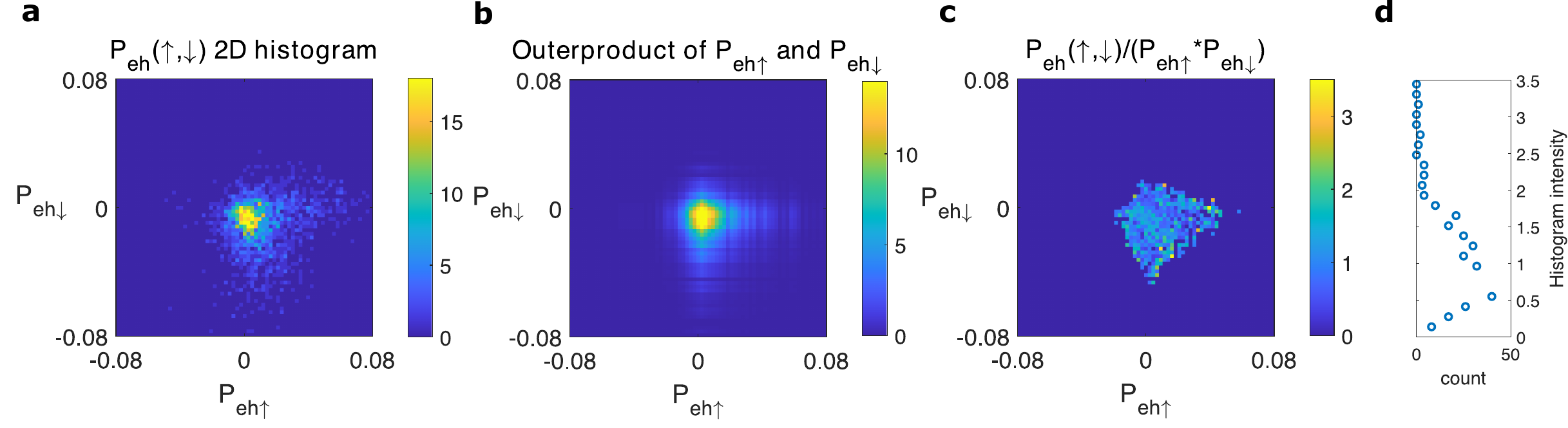}
    \caption {\label{fig:2Dhist}  a) Reproduced from Figure~3 of the main text: 2D histogram $P_\mathrm{eh}(\uparrow,\downarrow)$ with $P_\mathrm{eh}$ values extracted from the simultaneously measured $R_\mathrm{d\uparrow}$ and $R_\mathrm{d\downarrow}$. b) Outer product of $P_\mathrm{eh\uparrow}$ and $P_\mathrm{eh\downarrow}$ 1D histograms, normalized by their total count. This map corresponds to the expected $P_\mathrm{eh}(\uparrow,\downarrow)$ in case of uncorrelated spin-up and spin-down channels. Similarity between (a) and (b) is  visible. c) Ratio of the maps in (a) and (b). If $P_\mathrm{eh\uparrow}$ and $P_\mathrm{eh\downarrow}$ are not correlated, the ratio should be equal to one (see main text). Indeed, a plateau is seen in the central diamond-shaped region of the map. Here, regions outside  the diamond with counts less than 1 are masked out. d) Histogram of the map in (c), showing a broad distribution reasonably close to  1.
    }
\end{figure*}

\newpage
\subsection{Landauer-B\"uttiker formalism for different measurement setups}

Here we calculate the LB matrices for the sample shown in Figure~\ref{fig:optical} using the method of Ref.~\cite{Zhao2024}.  Only contact C is assumed to partially reflect the carriers.
In the case of filling factor $\nu=1$ in the bulk we obtain

\begin{equation}
\begin{pmatrix}
\delta I_A \\
\delta I_B \\
\delta I_C \\
\delta I_D \\
\delta I_E \\
\delta I_F \\
\delta I_G
\end{pmatrix}
=
\frac{e^2}{h}
\begin{pmatrix}
 g_1 & 0 & 0 & 0 & 0 & 0 & -g_1 \\
 -g_1 & g_1 & 0 & 0 & 0 & 0 & 0 \\
 0 & g_1(-1+P_{11}) & g_1(1-P_{11}) & 0 & 0 & 0 & 0 \\
 0 & -g_1 P_{11} & -g_1(1-P_{11}) & g_1 & 0 & 0 & 0 \\
 0 & 0 & 0 & -g_1 & g_1 & 0 & 0 \\
 0 & 0 & 0 & 0 & -g_1 & g_1 & 0 \\
 0 & 0 & 0 & 0 & 0 & -g_1 & g_1
\end{pmatrix}
\begin{pmatrix}
\delta V_A \\
\delta V_B \\
\delta V_C \\
\delta V_D \\
\delta V_E \\
\delta V_F \\
\delta V_G
\end{pmatrix},
\end{equation}
where we follow the notation of Ref.~\cite{Zhao2024} and denote the number of channels in the outer edge state as $g_1=1$. Next, we write the conductance matrix in $\nu=2$ case, where $g_2=1$ denotes the number of channels in the inner edge state:
\begin{equation}
\resizebox{\textwidth}{!}{$
\begin{pmatrix}
 g_1 & 0 & 0 & 0 & 0 & 0 & -g_1 \\
 -g_1 & g_1 & 0 & 0 & 0 & 0 & 0 \\
 0 & g_1(-1 + P_{11}) + g_2 P_{21} & g_1(1 - P_{11} - P_{12}) + g_2(1 - P_{21} - P_{22}) & 0 & 0 & 0 & g_1 P_{12} + g_2(-1 + P_{22}) \\
 0 & -g_1 P_{11} & -g_1(1 - P_{11} - P_{12}) & g_1 & 0 & 0 & -g_1 P_{12} \\
 0 & 0 & 0 & -g_1 & g_1 & 0 & 0 \\
 0 & -g_2 P_{21} & -g_2(1 - P_{21} - P_{22}) & 0 & -g_1 & g_1 + g_2 & -g_2 P_{22} \\
 0 & 0 & 0 & 0 & 0 & -(g_1 + g_2) & g_1 + g_2
\end{pmatrix}
$}
\end{equation}
%

Here the subscripts  of $P$ indicate the conversion rates for the (1) $\uparrow$ and (2) $\downarrow$ channels. For example, $P_{12}$
indicates the transition probability from the spin-up to the spin-down channel. Here we do not distinguish between electrons and holes, so both $e_\uparrow \rightarrow h_\downarrow$ Andreev reflection and $e_\uparrow \rightarrow e_\downarrow$ spin flip 
contribute to $P_{12}$.

We next use these equations to calculate the Andreev conversion rate for different setups in Figure~\ref{fig:SM_diffNotations}. 
In Figure~\ref{fig:SM_diffNotations}b, corresponding to biasing only spin-up electrons to the superconducting contact and measuring only the spin-up channel, the conversion from $R_\mathrm{d}$ to $P_{11}$ (spin-up to spin-up channel) is 
\begin{equation}
    R_\mathrm{d} \cong \frac{h}{e^2} \frac{2P_{11}}{2 - (P_1 + P_2)} 
    \qquad \qquad
    P_{11} \cong \frac{R_\mathrm{d}}{R_\mathrm{H(\nu=1)} + R_\mathrm{d}} \;.
\end{equation}
where we have assumed that $R_\mathrm{d}$ is much smaller than $R_\mathrm{H}$.

In Figure~\ref{fig:SM_diffNotations}c, corresponding to biasing only spin-down electrons to the superconductor and measuring only the spin-up channel, the conversion from $R_\mathrm{d}$ to $P_{21}$ (spin-down to spin-up channel) is
\begin{equation}
    R_\mathrm{d} = \frac{h}{e^2} \frac{P_{21}}{2 - P_2} 
    \qquad \qquad
    P_{21} \cong \frac{2R_\mathrm{d}}{R_\mathrm{H(\nu=1)} + R_\mathrm{d}} \;.
\end{equation}

In Figure~\ref{fig:SM_diffNotations}e, corresponding to only spin-up electrons entering the superconductor but measuring both spin channels, the conversion from $R_\mathrm{d}$ to $P_{1}$ is
\begin{equation}
    R_\mathrm{d} = \frac{h}{e^2} \frac{P_1}{2 - (P_1 + P_2)} 
    \qquad \qquad
    P_1 \cong \frac{R_\mathrm{d}}{R_\mathrm{H(\nu=2)} + R_\mathrm{d}}\;,
\end{equation}
where $P_{1}\equiv P_{11}+P_{21}$.

Finally, consider the case when the graphene is globally at $\nu=+1$, the current is biased from contact B, grounded at contact C (superconductor), and the voltage is measured at contact D.  It corresponds to only spin-up electrons entering the superconductor and measuring only the spin-up channel. The conversion from $R_\mathrm{d}$ to $P_{11}$ is
\begin{equation}
    R_\mathrm{d} = \frac{h}{e^2} \frac{P_{11}}{1 - P_{11}} 
    \qquad \qquad
    P_{11} = \frac{R_\mathrm{d}}{R_\mathrm{H(\nu=1)} + R_\mathrm{d}} \;.
\end{equation}

\newpage
\subsection{Verification of the  measurements' consistency with the  Landauer-B\"uttiker predictions for different biasing and detection setups}

\begin{figure*}[htp]
    \centering
    \includegraphics[width=0.9\columnwidth]{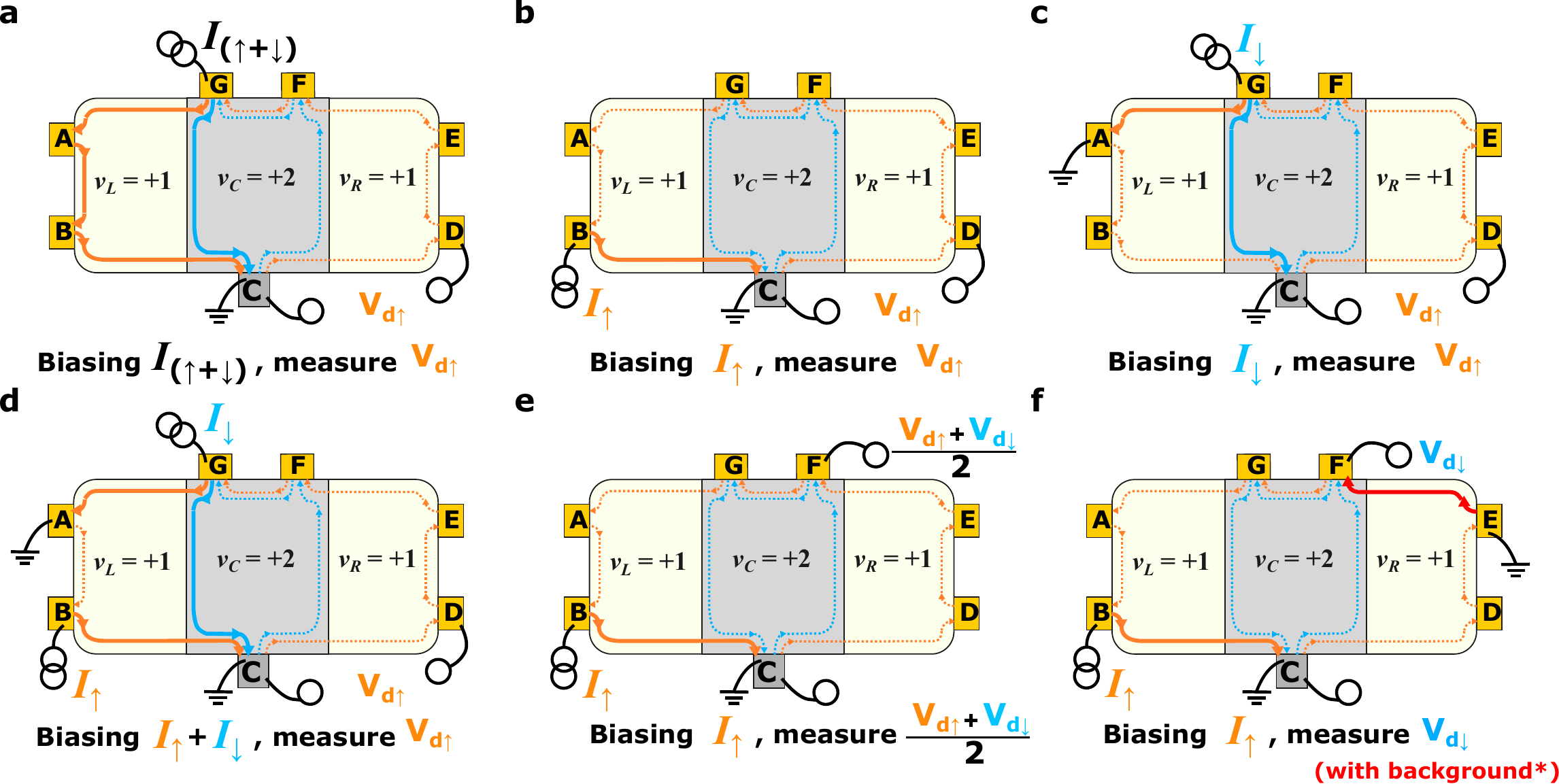}
    \caption {\label{fig:SM_diffNotations}    
    a-f) Different setups used in the measurement shown in the following figure. 
    }
\end{figure*}

\begin{figure*}[h!]
    \centering    \includegraphics[width=1\columnwidth]{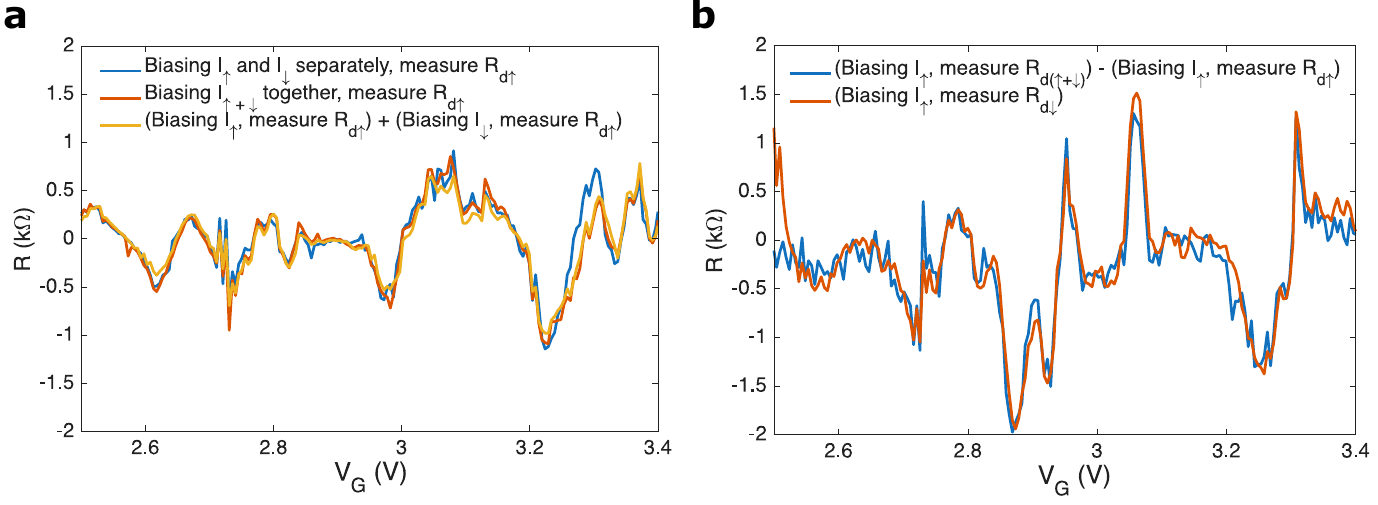}
    \caption {\label{fig:SM_inputoutput} $R_\mathrm{d}$ curves measured with different setups of Figure~\ref{fig:SM_diffNotations} to check the application of the LB formalism. (a) Spin-up $R_\mathrm{d\uparrow}$ curves measured with different setups. The red and blue curves correspond to setups (a) and (d) respectively. The yellow curve is an average of signals measured in setups (b) and (c). 
    b) Spin-down $R_\mathrm{d\downarrow}$ curves measured with different setups. The blue curve shows the result of subtracting the spin-up signal measured in setup (b) from the sum of both spin signals measured in setup (e). The red curve shows the spin-down signal measured in setup (f), subtracting the constant offset caused by double-grounding in the presence of cryogenic filters. 
    }
    
\end{figure*}

Figure~\ref{fig:SM_diffNotations} shows several measurement setups we have tested in Figure~\ref{fig:SM_inputoutput}. Below we summarize the properties of these setups and their relations.

Voltage probes: In setups (a-d), we probe the voltage at contact D, so that only the spin-up downstream signal is measured. Setup (e) measures both spin channels, while setup (f) measures only spin-down signal, because the spin-up channel is drained by contact E.  

Bias currents: In setup (a), contact G is biased, sending both spin currents. In setups (b), (e), and (f) we send only spin-up current from contact B. In setup (c) we send only spin-down current to contact C, since the spin-up current is drained at contact A. We denote half of the current applied to contact G as $I_\downarrow$.

Setup (d) is a combination of setups (b) and (c). Similarly to case (c),  half of the current applied to G is drained by contact A, while the remaining half $I_\downarrow$ flows to contact C. To send the same amount of $e_\downarrow$ and $e_\uparrow$ to contact C, we apply a twice smaller current to contact B as compared to contact G: $I_B=I_G/2$. 

Setup (f) is the only one that uses two grounds on the downstream side: contact E is grounded to drain the spin-up downstream current. Due to the presence of cryogenic filters, this scheme creates a constant offset at contact F that has to be subtracted from the measured $V_\downarrow$.

Finally, the reflected currents (dashed lines) can go all around the device and come back to contact C. However, they are negligible since the Andreev reflection rate ($P_\mathrm{eh}$) is small, and we neglect them in our considerations. 

\newpage

\newpage
\subsection{Random matrix theory with loss}

In this section, we present the random matrix theory (RMT) arguments
for the $S$-matrix that lead to Eqs.\,(2)-(3) of the main text. We
remind the reader that applying the $S$-matrix to the amplitudes in
the input channels yields the amplitudes in the output channels. Because
the scattering states are governed by the Bogoliubov-de Gennes (BdG)
equation describing mean-field superconductivity, there are an equal
number of electron and hole channels (this is the well-known doubling
of the degrees of freedom in the BdG description). The $S$-matrix must
be unitary and particle-hole symmetric. 

When the incoming and outgoing channels are quantum Hall edge states
at filling factor $\nu$, there are $\nu$ incoming electron channels
and an equal number of incoming hole channels, and likewise for the
outgoing channels. Thus in the absence of loss, the $S$-matrix has the
structure
\begin{align}
S_{\textrm{phys}} & =\begin{pmatrix}s_{ee} & s_{eh}\\
s_{he} & s_{hh}
\end{pmatrix},\quad s_{hh}=s_{ee}^{*},\;s_{he}=s_{eh}^{*},
\label{eq:Sphys}
\end{align}
where each block in the $S$-matrix is $\nu\times\nu$ (we assume that
all the electron channels are listed before their corresponding hole
channels). The given relations between the blocks are the necessary
and sufficient conditions for particle-hole symmetry \cite{BeenakkerRMP2015}. A system
may have additional symmetries, such as time-reversal, spin-rotation,
or parity symmetry, in which case the $S$-matrix is further constrained
 \cite{BeenakkerRMP2015}. 

Loss can be introduced by adding additional fictitious channels which
drain the electrons and holes away from the system \cite{ButtikerPRB86, BarangerMello-PhaseBreak-PRB95, BrouwerBeen-PhaseBreak-PRB95}. Perhaps
the best way to do this is to use a very large number of weakly coupled
channels~\cite{BrouwerBeen-PhaseBreak-PRB95}. However, when the loss is large, introducing channels
that are coupled as strongly as the physical channels yields the same
answer \cite{BarangerMello-PhaseBreak-PRB95, BrouwerBeen-PhaseBreak-PRB95} and is considerably simpler. Introducing $N_{\textrm{loss}}$
electron loss channels, the $S$-matrix becomes
\begin{align}
S_{\textrm{phys+loss}} & =\begin{pmatrix}s_{ee} & \textrm{e-eloss} & s_{eh} & \textrm{e-hloss}\\
\textrm{eloss-e} & \textrm{eloss-eloss} & \textrm{eloss-h} & \textrm{eloss-hloss}\\
s_{he} & ... & s_{hh} & ...\\
 & ... &  & \textrm{hloss-hloss}
\end{pmatrix},
\end{align}
where each quadrant of the $S$-matrix is now $(\nu+N_{\textrm{loss}})\times(\nu+N_{\textrm{loss}})$. 
Because the loss respects the particle-hole symmetry of the system (BdG description), if we define quadrants of this larger $S$-matrix,
\begin{align}
s_{ee,\textrm{loss}} & \equiv\begin{pmatrix}s_{ee} & \textrm{e-eloss}\\
\textrm{eloss-e} & \textrm{eloss-eloss}
\end{pmatrix},
\quad s_{hh,\textrm{loss}}=s_{ee,\textrm{loss}}^{*},\;s_{he,\textrm{loss}}=s_{eh,\textrm{loss}}^{*},
\end{align}
the same particle-hole symmetry relations hold between quadrants.

We suppose that the strong disorder in the system maximally mixes
the scattered waves in the scattering region and therefore consider
maximally random $S$-matrices as representative of the system. Thus,
we consider an ensemble of $S$-matrices that respect the required global
symmetries but are otherwise distributed uniformly across the matrix
space. For $S$-matrices with no restriction beyond unitarity, this is
known as the Haar measure for unitary matrices, and the ensemble is
known as the circular unitary ensemble (CUE). Additional symmetries
reduce the measure \cite{Beenakker-RMT-RMP97}. Our system has particle-hole symmetry, but no
additional symmetry because of the magnetic field, Zeeman splitting,
and spin-orbit (SO) scattering. In the classification of disordered superconducting
systems, this is called class D symmetry and corresponds
to the circular real ensemble (CRE) \cite{BeenakkerRMP2015}. 

For a large random $S$-matrix, the global symmetry and unitarity conditions
constrain an individual element of the matrix only weakly,
and the distribution of each matrix element will be the same (except
possibly for diagonal compared to off-diagonal elements). For instance
for large CUE matrices, the distribution of the magnitude of any given matrix element
is \cite{PereyraMello-MarginalJPA83}
\begin{align}
\mathcal{P}(|S_{ij}|) & =\frac{2}{\sqrt{\pi\alpha}} e^{-|S_{ij}|^{2}/\alpha},\quad\alpha=\frac{2}{N}\;.
\label{eq:UnitaryPSij}
\end{align}
The gaussian distribution here is, in fact, a general feature of the
circular ensembles \cite{Meckes-RMTbook2019}. We motivate this by, first, a semiclassical
central-limit-theorem argument and, second, a projection of measure
argument. 

In the semiclassical argument \cite{PolianskiBrouwerJPA03}, one considers the amplitude
along paths in position space that contribute to $S_{ij}$. Each $S_{ij}$
has contributions from many paths through the disordered scattering
region. By the central limit theorem, the sum of these random amplitudes
leads to a complex gaussian distribution for $S_{ij}$ and thus a
gaussian distribution for the amplitude. 

In the projection argument \cite{Meckes-RMTbook2019}, one first notes that in $S$-matrix
RMT, the measure in a large matrix space is uniform in a large but
compact (and convex) region. This region has an unintuitive shape because of the
unitarity and symmetry conditions. The probability density of $S_{ij}$
is obtained by projecting onto a single dimension. In the large matrix limit,
the projected density is rigorously known to be gaussian for several
circular ensembles (starting with Borel in 1906 for random unit vectors) and used heuristically in many other situations \cite{Meckes-RMTbook2019}. 

In our system, the electron-hole coupling is strong because of a transparent
interface; in contrast, the SO scattering does not fully mix
the two spin states. Thus, the measure for random $S$-matrices describing
our system is \textit{not} uniform. We assume that all matrix elements that
require a SO scattering event are statistically equivalent,
as are all those that do not. On the other hand, electron and hole
degrees of freedom are statistically equivalent. 

Both the semiclassical and projection arguments point to a gaussian
distribution for individual $S$-matrix amplitudes as neither argument
uses details of the measure. However, the distribution of matrix elements
that require SO scattering may be different from those that
do not. This implies that the coefficient $\alpha$ above in Eq.\,(\ref{eq:UnitaryPSij})
may be different for processes that require SO from those that do
not. Because of the requirement that $\mathrm{Tr}(SS^\dagger)=N$, all $\alpha$ must in any case be proportional to $1/N$.

We have thus established Eqs.\,(2) and (3) of the main text. 

We end with a brief comment about the \textit{lossless} $\nu=1$ case.  $S$ is $2\times2$, and the entries in Eq.\,(\ref{eq:Sphys}) are c-numbers. The three conditions on $S$ are
\begin{align}
S_{(\nu=1)} & =\begin{pmatrix}s_{ee} & s_{he}^*\\
s_{he} & s_{ee}^*
\end{pmatrix},\quad S_{(\nu=1)}^{\phantom{\dagger}} S_{(\nu=1)}^\dagger = I ,\quad \det\left(S_{(\nu=1)}\right)>0 \;.
\end{align}
Satisfying these conditions requires $|s_{he}|=0$.

\newpage
\subsection{Tight-binding calculation}

We implemented tight-binding transport calculations using Kwant~\cite{GrothKwantNJP14} by adapting the code from Ref.~\cite{Manesco2022, antonio_l_r_manesco_2021_4597080}, adding Zeeman splitting and spin-orbit coupling to the model.
Namely, we implemented the Hamiltonian
\begin{align}
  \label{eq:tb_hamiltonian}
  \mathcal{H} &= \sum_{i} \psi^{\dagger}_i (\Delta_i \tau_x - \mu_i \tau_z + E_Z\sigma_z) \psi_i - \sum_{\langle i, j \rangle} \psi_i^{\dagger} \tau_z \left[te^{i \tau_z \phi_{ij}}  +\alpha (\hat{r}_{ij}\times \hat{z}) \cdot \boldsymbol{\sigma}\right]\psi_j~,
\end{align}
where $\psi_i = (c_{i\uparrow},c_{i\downarrow},-c_{i, \downarrow}^{\dagger},c_{i\uparrow}^{\dagger})^T$, $c_{i\sigma}^{\dagger}$ and $c_{i\sigma}$ are the electron creation and annihilation operators at the position $\mathbf{r}_i$ and spin $\sigma$, and $\langle i,  j \rangle$ are all the pairs of nearest neighbor sites.
The Pauli matrices $\tau_i$ and $\sigma_i$ act on the electron-hole and spin degrees of freedom.
Using a honeycomb lattice, we simulate a device with a normal-superconductor interface located at $x=0$, using the following position dependence of the chemical potential $\mu_i$ and the superconducting pairing potential $\Delta_i$:
\begin{equation}\label{eq:spatial-dependency}
  \mu_i = (\mu_\mathrm{SC} - \mu_\mathrm{QH}) \Theta(x_i) + \mu_\mathrm{QH} + m_i \Theta(-x_i),\quad
  \Delta_i = \Delta \Theta(x_i)~,
\end{equation}
with $\mu_{QH}$ and $\mu_{SC}$ the onsite energies in the normal and the superconducting regions, and $m_i$ a sublattice-dependent staggered potential.
We add the orbital magnetic field contribution to the Hamiltonian via a Peierls phase:
\begin{align}
  \phi_{ij} = - \frac{\pi B}{\phi_0} (y_j - y_i) (x_j + x_i) \Theta\left(- \frac{x_i + x_j}{2} \right)~,
\end{align}
where $B$ is the orbital magnetic field, $\phi_0 = h / e$ is the magnetic flux quantum, and $\Theta(x)$ is the Heaviside step function.

To reduce the computational cost, we rescale the model $a \mapsto \tilde{a} = s a$ and $t \mapsto \tilde{t} = t / s$ to keep the Fermi velocity $v_F \propto ta$ unchanged.
In all transport calculations, we use $t = \qty{2.8}{\electronvolt}$, $a = \qty{0.142}{\nano\meter}$, $s = 10$, $B=\qty{1}{\tesla}$, $\mu_\mathrm{SC} = \tilde{t}/2$.
We choose $E_Z = 3\Delta / 4$ and $|m_i| = 3 E_Z$.
For this reason, we set $\mu_\mathrm{QH} = |m_i|$ at $\nu=1$, and $\mu_\mathrm{QH} = |m_i| + 2 E_Z$ at $\nu =2$.
Since Refs.~\cite{Manesco2022, Kurilovich2023} show that disorder in the superconductor plays a major role in the nonlocal transport, we add an uncorrelated disorder landscape to the Hamiltonian
\begin{align}
  \mathcal{H}_{\text{disorder}} = \sum_{i} \psi_i^{\dagger} \delta \mu_i\Theta(x_i) \tau_z \psi_i,
\end{align}
with $\delta \mu_i \in [-3\tilde{t}, 3\tilde{t}]$, resulting in a mean free path of $l_\mathrm{mfp} \approx \qty{13}{\nano\meter}$.
Finally, because the superconducting state only exists for $B < H_{c2}$, we conclude that the coherence length of the disordered superconductor is smaller than the magnetic length $l_B \approx \qty{26}{\nano \meter}$.
For this reason $\Delta = \qty{5}{\m\electronvolt}$, resulting in $\xi_0 = \hbar v_F / \Delta \approx \qty{38}{\nano\meter}$, resulting in a ``dirty'' coherence length $\xi = \sqrt{\xi_0 l_\mathrm{mfp}} = \qty{19}{\nano\meter}$.

We add loss to the simulations by creating a T-shaped device geometry, similarly to the one simulated in~\cite{Manesco2022} and connecting the superconducting region to a metallic reservoir.
We find that a superconducting region with width $W = \qty{25}{\nano\meter}$ reproduces the survival probabilities obtained experimentally.
These numbers are consistent with the typical distance of the closest vortices to the normal-superconductor interface being $\sim \xi$.

In the transport simulations, we compute the spin-resolved transmission rates, which include two possible processes: an incoming electron ($T_{ee}$) is transmitted as an electron or as a hole ($T_{he}$).
We then compute outcome probability $P_{eh} = T_{ee} - T_{he}$ for 2000 disorder configurations to build the histograms shown throughout the main text.

\newpage
\bibliographystyle{unsrturl}
\bibliography{Reference,QTransport_2026-07,RMT_gr-S}
